\documentclass[a4paper,fleqn,usenatbib]{mnras}

\usepackage{newtxtext,newtxmath}

\usepackage[T1]{fontenc}
\usepackage{ae,aecompl}
\usepackage{hyperref}
\usepackage{adjustbox}

\usepackage{graphicx}	
\usepackage{amsmath}	
\usepackage{amssymb}	

\newcommand{\vsini}{\ensuremath{{\upsilon}\sin i}}
\newcommand{\kms}{\,km\,s$^{-1}$}

\title{K2 Observations of 95 Vir: $\delta$ Scuti Pulsations in a Chromospherically Active Star}

\author[E. Paunzen et al.]{
Ernst Paunzen,$^{1}$\thanks{E-mail: epaunzen@physics.muni.cz}
Stefan H{\"u}mmerich,$^{2,3}$
Klaus Bernhard,$^{2,3}$
Przemek Walczak$^{4}$
\\
$^{1}$Department of Theoretical Physics and Astrophysics, Masaryk University, Kotl\'a\v{r}sk\'a 2, 611 37 Brno, Czech Republic\\
$^{2}$American Association of Variable Star Observers (AAVSO), Cambridge, USA\\
$^{3}$Bundesdeutsche Arbeitsgemeinschaft f{\"u}r Ver{\"a}nderliche Sterne e.V. (BAV), Berlin, Germany\\
$^{4}$Instytut Astronomiczny, Uniwersytet Wroc{\l}awski, PL-51-622 Wroc{\l}aw, Poland\\
}

\date{Accepted XXX. Received YYY; in original form ZZZ}

\pubyear{2017}

\begin{document}
\label{firstpage}
\pagerange{\pageref{firstpage}--\pageref{lastpage}}
\maketitle


\begin{abstract}
We have searched for photometric variability in 95 Vir, a fast rotating, chromospherically active early F-type star, which was observed in the framework of Campaign 6 of the Kepler K2 mission. Available literature information on 95 Vir were procured, and well-established calibrations were employed to verify the derived astrophysical parameters. We have investigated the location of our target star in the $M_{\rm Bol}$ versus $\log T_\mathrm{eff}$ diagram, which provides information on evolutionary status. We have discussed our results in detail, drawing on literature information and the theoretical predictions of state-of-the-art pulsation models, with the aim of unraveling the underlying variability mechanisms. From an analysis of 3400 long-cadence measurements, we have identified two main frequencies and several harmonics in our target star. We attribute the main frequency, $f1$\,=\,9.53728\,d$^{-1}$, to $\delta$ Scuti pulsations. The origin of the secondary signal, $f2$\,=\,1.07129\,d$^{-1}$, is less clear. We have investigated three possible interpretations of the low-frequency variation: binarity, pulsation and rotational modulation. Current evidence favours an interpretation of $f2$ as a signature of the rotational period caused by the presence of cool star spots, which goes along well with the observed chromospheric activity. However, phase-resolved spectroscopy is needed to verify this assumption. We briefly consider other chromospherically active $\delta$ Scuti stars that have been presented in the literature. A search for star spot-induced photometric variability in these objects might be of great interest, as well as an investigation of the interplay between chromospheric and pulsational activity.
\end{abstract}

\begin{keywords}
stars: activity -- stars: variables: delta Scuti -- stars: individual: HD 123255
\end{keywords}



\section{Introduction} \label{introduction}

$\delta$ Scuti (GCVS-type DSCT) stars have long been established as a class of variable stars \citep{fath35}. They are multiperiodic pulsators of luminosity classes V to III that boast masses between about 1.5\,$\rm\,M_\odot$\,<\,$M$\,<\,4.0\,$\rm\,M_\odot$ \citep{aerts10}. According to \citet{catelan15} and the General Catalog of Variable Stars (GCVS; \citealt{gcvs}), they are mostly found among spectral types A0 to F5. Other sources \citep[e.g.][]{breger98} have located the blue border of the $\delta$\,Scuti instability strip at cooler temperatures (spectral type of $\sim$A2).

$\delta$ Scuti stars typically show variability on timescales between about 15 minutes and 5 hours \citep{breger00,gcvs,holdsworth14}. \citet{grigahcene10} defined $\delta$ Scuti stars as A-F type pulsators exhibiting frequencies in the domain $f$\,>\,5\,d$^{-1}$. The observed light changes are caused by multiple radial and non-radial low-order pressure (p) modes, which are excited through the $\kappa$ mechanism \citep[e.g.][]{breger00,handler09}. In evolved $\delta$ Scuti stars, so-called mixed modes are often observed. These are pulsation modes exhibiting gravity (g) mode characteristics in the interior and p mode characteristics near the stellar surface \citep[e.g.][]{lenz10,bowman16}. The occurence of mixed modes renders $\delta$ Scuti stars prime targets for asteroseismological studies because they probe different layers inside a star \citep{lenz10}. In general, hot $\delta$ Scuti stars have shorter periods than their cooler counterparts.

In some $\delta$ Scuti stars, only one or two radial modes are excited (usually the fundamental mode and/or first harmonic) and the observed peak-to-peak amplitudes exceed 0.3\,mag. These stars are known as High Amplitude Delta Scuti (HADS) variables \citep{mcnamara00}. HADS stars are usually slow rotators, which seems to be a requirement for the observed high-amplitude pulsation. Apart from this particular group of stars, $\delta$ Scuti variables are generally moderate to fast rotators \citep{breger00}.

Despite a lot of observational and theoretical effort, $\delta$ Scuti stars are still not well understood. Several open questions remain, e.g. regarding the relationship of $\delta$ Scuti stars to the class of $\gamma$ Doradus variables, whose theoretical instability regions on the Hertzsprung-Russell diagram overlap \citep{dupret04}, or the variable pulsation amplitudes observed in these stars \citep{bowman16}. It has been shown that many DSCT stars are in fact hybrid pulsators, exhibiting both p and gravity (g) modes \citep[e.g.][]{uytterhoeven11}. Generally, $\delta$ Scuti stars often exhibit low-frequency peaks in their frequency spectra. However, it is open to debate whether these frequencies are due to pulsation, the effects of rotation or some other cause, e.g. combination frequencies of high-frequency pulsation modes \citep[e.g.][]{bowman16}.

Chromospheric activity is a ubiquitous feature of lower main-sequence stars with convective envelopes below their photospheres \citep{wilson78}. Diverse phenomena are lumped together under this heading, such as e.g. the occurrence of starspots and magnetic activity cycles. It has been shown that chromospheric activity is related to the stellar rotation rate and the properties of the convection zones needed to generate the observed magnetic fields \citep[e.g.][]{noyes84}.

Convection zones appear in stars of spectral type $\sim$F0 ($T_\mathrm{eff}$ $\approx$ 7500 K) and deepen with decreasing temperature \citep{gilliland85}. However, the precise effective temperature or spectral type where chromospheric activity sets in is uncertain \citep[e.g.][]{walter95}. While virtually all main-sequence stars with spectral types later than F0 show emission from the transition and coronal regions \citep{wolff87}, significant chromospheric emission seems to be present in at least some stars as hot as $\sim$8000 K \citep{walter95}. Unfortunately, standard chromospheric activity indicators become more difficult to observe in late A / early F-type stars, e.g. due to rotational broadening of the spectral lines \citep{rachford09}.

Many surveys contributed to the currently available extensive pool of chromospheric activity measurements \citep[e.g.][]{henry96,wright04}. In general, chromospheric activity is measured by determining the flux in $\sim$1\,\AA\,wide filters centered on the \ion{Ca}{ii} H\&K lines, which is then normalized to the flux in two continuum filters placed short- and longward of the \ion{Ca}{ii} H\&K region \citep{schroeder09}. Obviously, rotational line broadening interferes with these measurements, so that this method can be applied to slow rotators only. However, it has been shown that this obstacle can be overcome by e.g. the fitting of artificially broadened template spectra to a rapid rotator's line-wing spectrum \citep{schroeder09}.

The surfaces of most chromospherically active stars are covered with spots, which results in photometric variability with the rotation period. In this way, precise rotation periods can be determined.

Several studies have investigated the occurrence of chromospheric activity in $\delta$ Scuti stars \citep[e.g.][]{fracassini82,Teays1989,fracassini91}. Evident chromospheric emissions were identified in some $\delta$ Scuti stars, and the existence of a relation between chromospheric emission widths and pulsational periods was suggested \citep{fracassini82}. However, as far as we are aware of, this question has not been tackled in detail yet. To lay the ground for further studies, it is important to increase the sample size of chromospherically active $\delta$ Scuti stars.

\begin{table}
\caption{Astrophysical parameters of 95 Vir from the literature.}
\label{table_astro}
\begin{center}
\begin{adjustbox}{max width=0.5\textwidth}
\begin{tabular}{p{6mm}cccccc}
\hline
\hline 
$T_\mathrm{eff}$ & $\log g$ & Age & [Fe/H] & Mass & \vsini\ & Ref \\
(K) & (dex) & (Gyr) & (dex) & ($\rm M_\odot$) & (\kms) & \\
\hline
6792 & 3.78	& & $-$0.14	& & & 1 \\
6918 & 3.63 & & & & & 2 \\
6886 & & & & & & 3 \\
7031 & 3.85 & &	& 1.91 & 180 & 4 \\		
7014 & & & & & 157.8 & 5 \\
& & 1.50 & $-$0.04 & & & 6 \\ 
& & 1.25 & & & & 7 \\
& & & & 1.75 & & 8 \\
& & & & & 140 & 9 \\
\hline
\multicolumn{7}{l}{{1 ... \citet{Erspamer2003},
2 ... \citet{schroeder09}}} \\
\multicolumn{7}{l}{{3 ... \citet{Casagrande2011},
4 ... \citet{Zorec2012}}} \\
\multicolumn{7}{l}{{5 ... \citet{Ammler2012},
6 ... \citet{Gontcharov2012}}} \\
\multicolumn{7}{l}{{7 ... \citet{Pace2013},
8 ... \citet{prieto99}}} \\
\multicolumn{7}{l}{{9 ... \citet{vanbelle12}}}
\end{tabular}
\end{adjustbox}
\end{center}
\end{table}

\section{Target star, literature information and astrophysical parameters} \label{targetstar}

The star 95 Vir (HD 123255, HR 5290, $V$\,=\,5.47\,mag) has been very well studied. \citet{Cowley1974} classified it as F2\,IV whereas \citet{Gray1989} list a spectral type of F0\,IV (Cr). Several studies have determined our target star's astrophysical parameters; an overview of these results is presented in Table \ref{table_astro}. From this table, we deduce the following values for 95 Vir, an object of solar abundance: $T_\mathrm{eff}$\,=\,6928(98)\,K and $\log g$\,=\,3.75(11)\,dex. To verify these values, we used Str\"omgren-Crawford $uvbyH\beta$ and $Geneva$ colours from the ``The General Catalogue of Photometric Data'' (GCPD\footnote{http://gcpd.physics.muni.cz/}) and \citet{Paunzen2015} as well as the calibrations by \citet{Napiwotzki1993} and \citet{Kunzli1997}. The obtained values are well in line with the above-listed deductions and have been employed in the construction of a $M_{\rm Bol}$ versus $\log T_\mathrm{eff}$ diagram, as discussed below.

The classification of a star as luminosity class IV (subgiant) implies that it has reached the terminal-age main sequence (TAMS) and established a hydrogen-burning shell source around a helium core. However, on the basis of \textsc{Hipparcos} data, \citet[][]{Paunzen1999} pointed out the controversial status of luminosity class IV \citep[][]{Keenan1985} and found no distinction in absolute luminosity between stars of spectroscopic luminosity classes IV and V. Furthermore, there exists a considerable overlap between these classes and luminosity class III, with only the brightest class III objects being well-separated from those of class V. This implies that luminosity class IV should
be rejected.

We have investigated the location of 95 Vir within the $M_{\rm Bol}$ versus $\log T_\mathrm{eff}$ diagram, which provides information on evolutionary status. As listed in Table \ref{table_astro}, published ages vary between 1.25 and 1.50\,Gyr. The calibration within the PARSEC isochrones \citep{Bressan2012} for solar metallicity ([Z]\,=\,0.019\,dex) yields an almost perfect match with the evolutionary track of $\log t$\,=\,9.1 (1.26\,Gyr), which implies that 95 Vir is still on the main sequence and therefore an object of luminosity class V. This finding is in line with the result by \citet{Zorec2012}, who published a fractional age of 0.881(16).

In order to look for a possible infrared excess, we have fitted the spectral energy distribution (SED) of 95 Vir using the VO Sed Analyzer \citep[VOSA;][]{Bayo2008} tool. All available photometric data were procured and included into the fitting process. No IR-Excess up to 25$\mu$m was detected; redwards of this limit, only upper limits for fluxes are available.

\begin{figure}
\begin{center}
\includegraphics[width=0.40\textwidth]{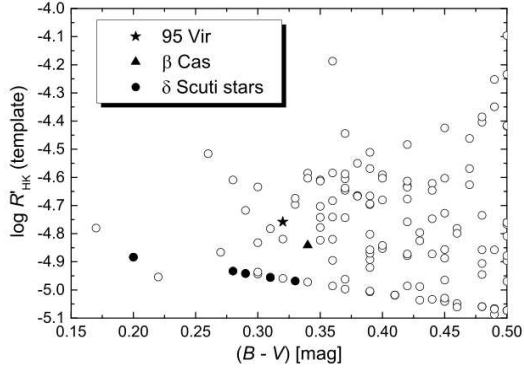}
\caption{The color-independent chromospheric activity index $\log R'_\mathrm{HK}$ versus Johnson $(B - V)$ index for stars cooler than late A-type from \citet{schroeder09}. Filled symbols indicate known $\delta$ Scuti pulsators. The chromospherically active stars 95 Vir and $\beta$ Cas are marked by different symbols, as indicated in the inset legend.}
\label{activity}
\end{center}
\end{figure}

During an investigation of chromospheric \ion{Ca}{ii} H\&K emission in rapidly rotating stars, \citet{schroeder09} searched for signs of chromospheric activity in 95 Vir (cf. Fig. \ref{activity}). As has been pointed out above (cf. Section \ref{introduction}), activity is measured photometrically and spectroscopically through the \ion{Ca}{ii} H\&K lines. \citet{Vaughan1978} introduced the dimensionless $S$-index as an indicator for the \ion{Ca}{ii} activity, which depends on the colour of the star and both chromospheric and photospheric radiation \citep{henry96}. To remove the colour dependence and the photospheric component, \citet{Middelkoop1982} and \citet{noyes84} developed a transformation of the $S$-index into a value $R'_\mathrm{HK}$ as a function of $(B - V)$. The corresponding values for 95 Vir are $S$\,=\,0.22 and $\log R'_\mathrm{HK}$\,=\,$-$4.758. According to \citet{henry96}, our target belongs to the `active' class, which is also obvious from Fig. \ref{activity}.

\section{Light curve analysis}

The following sections provide an overview over our data source and the employed methods of period analysis.

\subsection{Kepler K2 observations}

95 Vir was observed during Campaign 6 of the Kepler K2 mission during a time span of $\sim$78.9 days (July 13 to September 30, 2015). About 3400 measurements were collected in the long (29.4 min sampling) cadence mode. Since the loss of the second of its four reaction wheels, Kepler's onboard thrusters are used in regular intervals ($\sim$6 hours) to adjust the pointing of the spacecraft. This leads to the presence of characteristic systematics in the data sets \citep{howell14}, and care has to be taken in separating the astrophysical variability in K2 light curves from instrumental artefacts.

\begin{figure}
\begin{center}
\includegraphics[width=0.47\textwidth]{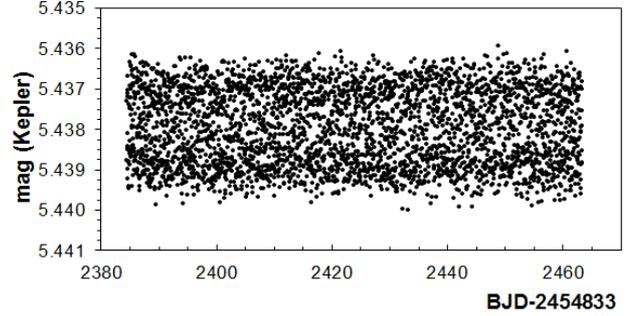}
\caption{Light curve of 95 Vir, based on Kepler K2 observations from the MAST archive. Obvious outliers have been removed by visual inspection.}
\label{lightcurve}
\end{center}
\end{figure}

We have downloaded the Kepler K2 light curve of 95 Vir from the archive of K2 Data Products at the Mikulski Archive for Space Telescopes (MAST).\footnote{https://archive.stsci.edu/k2/} Obvious outliers were removed by visual inspection. No obvious long-term trends of instrumental origin could be identified in this data set, which is of excellent quality. The K2 light curve of 95 Vir is shown in Fig. \ref {lightcurve}.

\subsection{Period analysis} \label{section_periodanalysis}

\begin{figure}
\begin{center}
\includegraphics[width=0.47\textwidth]{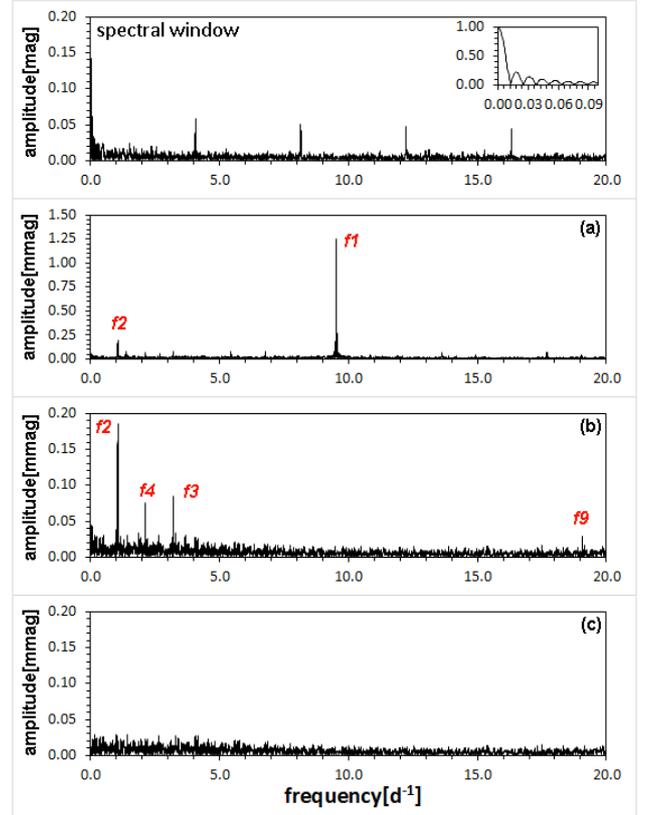}
\caption{Period analysis of K2 data for 95 Vir, illustrating the different steps of the frequency spectrum analysis. The top panel shows the spectral window dominated by signals at integers of $\sim$4.08\,d$^{-1}$, caused by the thruster firings that occur about every six hours. The inset shows a detailed view of the frequency region from 0.0 to 0.1 d$^{-1}$. The other panels illustrate the frequency spectra for (a) unwhitened data, (b) data that has been prewhitened with $f1$ and (c) the residuals after prewhitening with frequencies $f1$ to $f10$. Note the different scales on the ordinates.}
\label{periodanalysis}
\end{center}
\end{figure}

For the period analysis, we have employed the software package \textsc{Period04} \citep{period04}. In order to extract all relevant frequencies, the data were searched for periodic signals and consecutively prewhitened with the most significant frequency.  Periodograms and phased light curves were visually inspected in order to prevent instrumental signals from contaminating our results. K2 data for 95 Vir are very homogeneous and do not suffer from strong systematics, which renders the identification of the astrophysical signal a rather straightforward task. Nevertheless, signals at integers of $\sim$4.08\,d$^{-1}$ are present in the data set (cf. Fig. \ref{periodanalysis}, upper panel), which are caused by the thruster firings that occur about every six hours. These, however, are of comparably low amplitude in this data set and were further reduced in amplitude by careful removal of outlying datapoints. Our results are presented in Table \ref{pa_table} and Fig. \ref{periodanalysis}.

\begin{table}
\begin{center}
\caption{Frequencies, semi-amplitudes and signal-to-noise ratios detected in the Kepler K2 data for 95 Vir, as derived with PERIOD04. Signals attributed to intrinsic stellar variability are highlighted in bold type.}
\label{pa_table}
\begin{tabular}{ccccc}
\hline
\hline
ID & frequency  & semi-amp. & S/N & remark\\
   & [d$^{-1}$] & [mmag] & & \\       
\hline
$f1$	& \textbf{9.53728} & 1.250 & 221.65 & \\
$f2$	& \textbf{1.07129} & 0.188 & 16.23 & \\
$f3$	& \textbf{3.21707} & 0.085 & 8.56 & 3*f2 \\
$f4$	& \textbf{2.14327} & 0.077 & 7.74 & 2*f2 \\
$f5$	& 0.03311 & 0.044 & 3.87 & \\
$f6$	& 0.07613 & 0.044 & 4.49 & \\
$f7$	& 1.88808 & 0.033 & 3.37 & \\
$f8$	& 3.68246 & 0.031 & 2.94 & \\
$f9$	& \textbf{19.07520} & 0.030 & 5.91 & 2*f1 \\
$f10$	& 1.42938 & 0.029 & 3.31 & \\
\hline
\end{tabular}
\end{center}
\end{table}


\begin{figure}
\begin{center}
\includegraphics[width=0.47\textwidth]{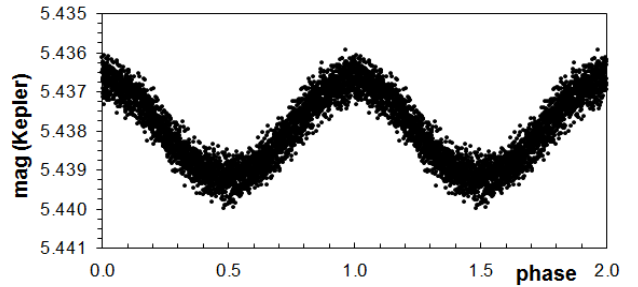}
\caption{Phase plot of 95 Vir, based on Kepler K2 observations and folded with the main frequency of $f1$\,=\,9.53728\,d$^{-1}$.} 
\label{phaseplot}
\end{center}
\end{figure}

\begin{figure}
\begin{center}
\includegraphics[width=0.47\textwidth]{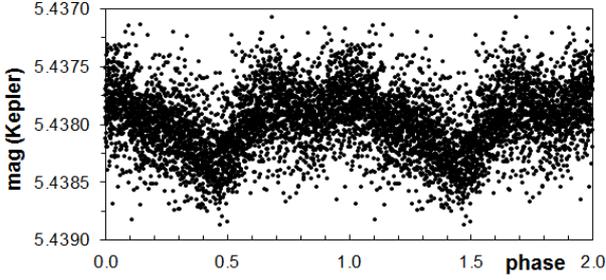}
\caption{Phase plot of 95 Vir, based on Kepler K2 observations that have been prewhitened for $f1$ and folded with $f2$\,=\,1.07129\,d$^{-1}$.}
\label{phaseplot2}
\end{center}
\end{figure}

K2 data for 95 Vir are dominated by a strong signal at $f1$\,=\,9.53728\,d$^{-1}$, boasting a S/N ratio in excess of 200 and a peak-to-peak amplitude of 2.5\,mmag. The phase plot folded with this signal is illustrated in Fig. \ref{phaseplot} and shows a near-sinusoidal light curve. After prewhitening for this frequency, we have derived $f2$\,=\,1.07129\,d$^{-1}$ and, subsequently, the corresponding harmonics $f3$\,=\,3.21707\,d$^{-1}$ (3*$f2$) and $f4$\,=\,2.14327\,d$^{-1}$ (2*$f2$). The phase plot folded with $f2$ is illustrated in Fig. \ref{phaseplot2} and displays a highly asymmetric and double-humped light curve. 

The low-S/N peaks $f5$ to $f8$, $f10$ are of unclear origin. The low frequencies $f5$\,$\approx$\,0.033\,d$^{-1}$ and $f6$\,$\approx$\,0.076\,d$^{-1}$ are likely due to the intrinsic characteristics of the data set. The presence of significant low-frequency power in K2 data, which is unrelated to intrinsic stellar variability, has been established \citep[e.g.][]{douglas16,gonzalez16}. Bearing in mind the time baseline of only $\sim$79 days, we are therefore inclined to regard these frequencies as spurious detections.\footnote{In this respect, it is interesting to note that $f6$ nearly exactly coincides with a harmonic of the campaign length.} The same conclusion holds true for the frequencies $f7$, $f8$ and $f10$ that boast S/N ratios well below 4, which is generally regarded as detection criterion \citep{breger93}. We therefore regard these frequencies as spurious detections.

The peak at $f9$\,=\,19.07520\,d$^{-1}$ (S/N$\sim$6) is obviously the second harmonic of the main frequency $f1$ (2*$f1$). Although this mode's peak-to-peak amplitude is only $\sim$60 $\mu$mag, it is a clear detection. We estimate the grass level in this frequency region to be less than $\sim$15 $\mu$mag, which underlines the high quality of the K2 data, which, at least in the case of the analyzed data set, are comparable in this respect to data from the original Kepler mission. Note, though, that the noise level increases significantly towards lower frequencies, which substantiates the above mentioned results of \citet{douglas16} and \citet{gonzalez16}.

It is interesting to note that, in the framework of a search for $\gamma$ Doradus pulsation in A- and F-type stars, \citet{Henry2011} presented photoelectric photometry of 95 Vir. From 36 observations taken over a timespan of almost 10 hours, they concluded that the star is constant with an intrinsic brightness variability less than 1.2\,mmag. This result, however, is in conflict with our finding of photometric variations with an amplitude of 2.5\,mmag.

\section{Discussion}

The dominant signal in 95 Vir, $f1$\,=\,9.53728\,d$^{-1}$, is well inside the frequency realm of the $\delta$ Scuti stars. Furthermore, 95 Vir is situated in the classical $\delta$ Scuti instability strip, and the near-sinusoidal light curve shown in Fig. \ref{phaseplot} is typical of this group of variable stars. Evidence therefore favours an interpretation as pulsational variability of $\delta$ Scuti-type.

The interpretation of the secondary frequency, $f2$\,=\,1.07129\,d$^{-1}$, is not as straightforward. As has been shown above, the phase plot folded on this frequency exhibits a quite peculiar, asymmetric and double-humped light curve (cf. Fig. \ref{phaseplot2}), which provides evidence against pulsational variability as the underlying cause. In the plot, phase $\varphi$\,=\,1.0 roughly denotes the time of maximum brightness. The drops in brightness at phases $\varphi$\,$\approx$\,0.5 and $\varphi$\,$\approx$\,0.85 are reminiscent of the minima observed in the light curves of eclipsing binary systems. Furthermore, the general light curve pattern vaguely resembles that of close binary systems, like e.g. W UMa-type variables, which are contact systems made up of similar stars generally belonging to spectral types F to K \citep{rucinski97}, and exhibit periods on the order of 0.2 to 1.0 days. In these close systems, both components are ellipsoidal and share a common envelope, which results in both stars sharing similar surface temperatures although the masses may be different \citep{jiang09}. However, these close binaries and related types are mostly synchronized and circularized systems. It is therefore hard to reconcile the observed asymmetric light curve with this scenario, which would imply a highly eccentric orbit. Furthermore, it is not clear how the dominant, short-period pulsational variability observed in 95 Vir fits into this picture. $\delta$ Scuti pulsations are not expected in W UMa stars; in any case, the proximity effects expected in the light curves of such close systems should exceed in amplitude a proposed $\delta$ Scuti variability of one of the components.

In fast rotating stars, the photosphere is distorted into an oblate spheroid by centrifugal forces. \citet{vanbelle12} estimated a rotationally-induced oblateness of 0.09 for 95 Vir. Photometric variability may be caused by the continuously changing aspect along the line-of-sight, induced by, for example, the gravitational pull of a companion star. Once again, however, (near) sinusoidal light variability would be expected in such a scenario, which is not in agreement with the observed variability in 95 Vir.

95 Vir has been the target of several spectroscopic studies (cf. Sect. \ref{targetstar}), in none of which any hint of duplicity was identified in the investigated spectra. In addition, the star's position in the $M_{\rm Bol}$ versus $\log T_\mathrm{eff}$ diagram is not indicative of any excess luminosity due to a close companion. To sum up, the assumption of a close binary system seems unsuited to explain the observed $\sim$0.93\,d period in our target star.

On the other hand, the observed asymmetric light curve shown in Fig. \ref{phaseplot2} is in general agreement with rotational variability caused by spots. This assumption is further corroborated by the presence of strong harmonics in the frequency spectrum (cf. Sect. \ref{section_periodanalysis}). Harmonics of pulsation modes are only expected in frequency spectra when the pulsational amplitude is large (as is the case with the main frequency $f1$ in 95 Vir, for which the second harmonic has been identified). On the other hand, harmonics are a consequence of localized spots and a characteristic of the frequency spectra of rotating variables \citep[e.g.][]{balona15}.

Although the star has been identified as exhibiting a possible Cr abundance peculiarity \citep{Gray1989}, we exclude the possibility of photometrically active abundance spots on grounds of the observed high rotational velocity (cf. Table 1). The strong rotationally-induced mixing should effectively prevent the formation of such features, which are a characteristic of the magnetic chemically peculiar stars \citep{mikulasek10}.

The assumption of the presence of solar-like starspots, on the other hand, goes along well with the observed chromospheric activity in 95 Vir (cf. Sect. \ref{targetstar}). Cool star spots, which result in photometric variability with the rotational period, are a ubiquitous feature of active stars. If we interpret $f2$ in this vein, the observed photometric period ($P$\,$\approx$\,0.93\,d) would be the rotational period. Assuming \vsini\ = 157.8\kms, which is at the low end of the available literature values (cf. Table \ref{table_astro}), \citet{Ammler2012} list a projected rotational period of 0.8 d, which is in general agreement with the above listed value. Adopting higher literature values of \vsini, as e.g. 180\kms\ as given by \citet{Zorec2012}, results in an even better agreement.

\begin{figure}
\begin{center}
\includegraphics[width=0.47\textwidth]{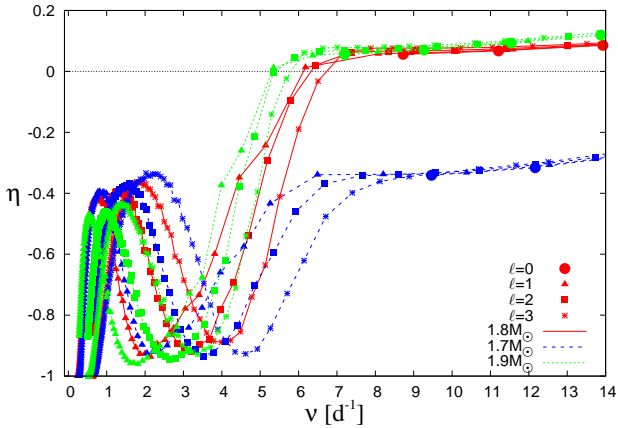}
\caption{The instability parameter $\eta$ as a function of frequency $\nu$ for 95 Vir. The pulsational stability properties for different mass regimes and modes of different degree ($\ell$\,=\,0,\,1,\,2 and 3) are shown (as indicated in the legend), assuming [Z]\,=\,0.015\,dex. See text for details.}
\label{pulsationmodel}
\end{center}
\end{figure}

In order to investigate our assumptions, in particular whether the low frequency $f2$ might be the result of an excited g mode, as is commonly observed in hybrid pulsators (cf. Sect. \ref{introduction}), we have investigated the predictions of pulsation models. Theoretical tracks were calculated for several different values of mass and metallicity. We have employed the new OPLIB opacity data \citep{colgan15,colgan16}, the scaled chemical composition by \citet{Aspl09}, the MESA code \citep{paxton11,paxton13,paxton15}, and the linear non-adiabatic pulsation code by \citet{Dzie77}. Additional models were calculated using the Opacity Project data \citep{Seat05}; the results were qualitatively the same.

Fig. \ref{pulsationmodel} illustrates the instability parameter $\eta$ as a function of frequency $\nu$ (d$^{-1}$); modes are excited when $\eta$\,>\,0. We have investigated the pulsational stability properties for different mass regimes and modes of different degree ($\ell$\,=\,0,\,1,\,2 and 3). Pulsations in the $\delta$ Scuti frequency realm are excited for models with masses of M > 1.7 $\rm M_\odot$, but the low-frequency modes are stable.

\begin{figure}
\begin{center}
\includegraphics[width=0.47\textwidth]{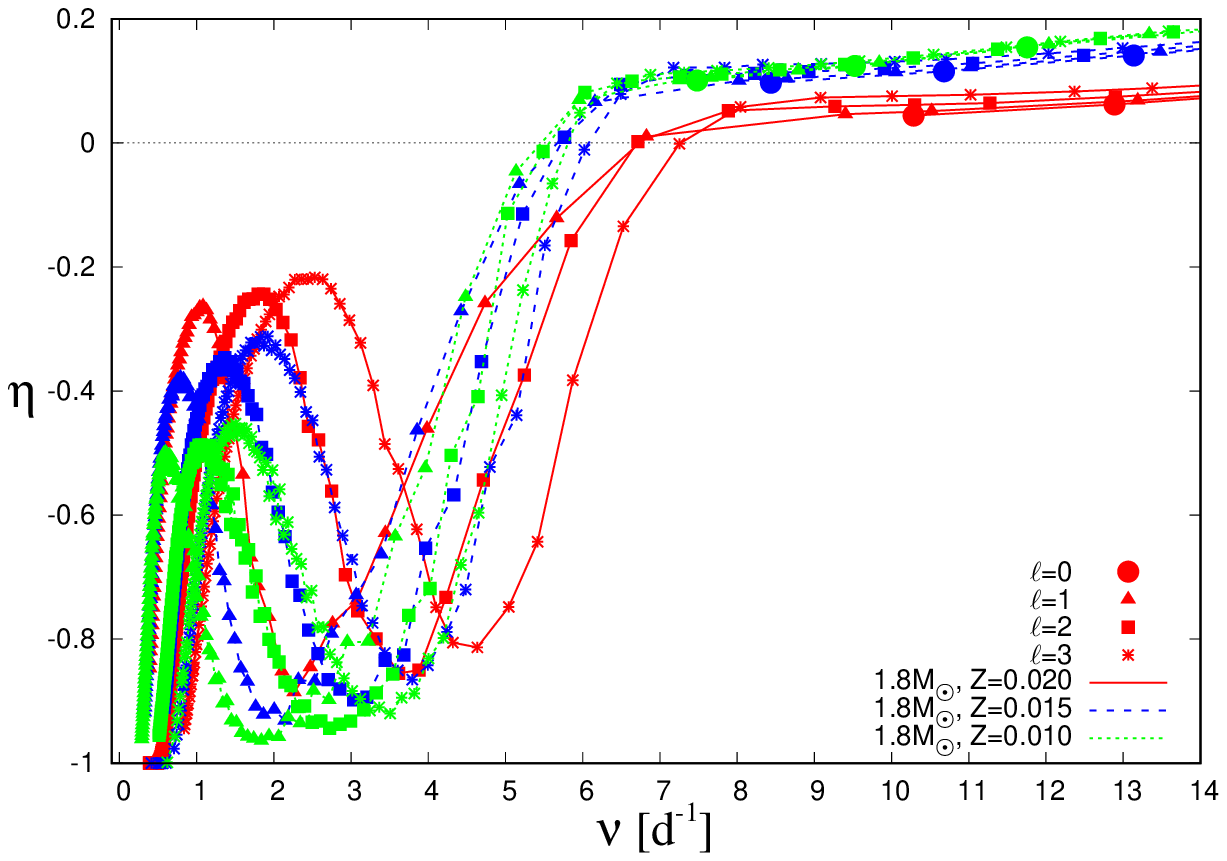}
\caption{The instability parameter $\eta$ as a function of frequency $\nu$ for 95 Vir. The pulsational stability properties for different metallicities and modes of different degree ($\ell$\,=\,0,\,1,\,2 and 3) are shown (as indicated in the legend), assuming M\,=\,1.8\,$\rm M_\odot$. See text for details.}
\label{pulsationmodel2}
\end{center}
\end{figure}

The model shown in Fig. \ref{pulsationmodel} has been calculated assuming a metallicity of [Z]\,=\,0.015\,dex. Fig. \ref{pulsationmodel2} investigates the influence of different metallicities on the predicted pulsational frequencies for a mass value of 1.8\,$\rm M_\odot$. While metallicity has quite a profound impact, the general picture does not change: no low-frequency modes are excited for any model. This provides more evidence against pulsational variability as the underlying mechanism for the observed $\sim$0.93 d variability in 95 Vir. It is worth to point out, though, that another source besides the classical $\kappa$ mechanism has been suggested for driving the g mode pulsations seen in the $\gamma$ Dor variables, namely the convective flux blocking mechanism \citep{guzik00}.

To sum up, based on the above-listed results and arguments, we interpret the secondary frequency $f2$\,=\,1.07129\,d$^{-1}$ ($P$\,$\approx$\,0.93\,d) in 95 Vir as a signature of the rotational period caused by the presence of cool star spots. However, phase-resolved spectroscopy is needed to verify this assumption, which we herewith encourage.

The question arises whether similar photometric variability can be established for other $\delta$ Scuti stars as well. As has been pointed out above (cf. Section \ref{introduction}), chromospheric activity has been reported in several other $\delta$ Scuti variables \citep[e.g.][]{fracassini82,Teays1989,fracassini91}. Another apparently chromospherically active $\delta$ Scuti star, $\beta$ Cas (HD 432, HR 21), is evident from Fig. \ref{activity}. \citet{Teays1989} analyzed the chromosphere of this star using spectra in the ultraviolet (UV) and visible regions as well as photometry. They found that the activity is modulated by the pulsations but that the observed mean level is comparable to other early F-type dwarfs. No attempt was made to search for rotationally-induced variability due to stellar spots.

The long-term photometric behaviour of $\beta$ Cas was analyzed by \citet{Riboni1994}, who concentrated on the $\delta$ Scuti pulsations (main frequency of $f$\,=\,9.899\,d$^{-1}$). The data sets of several consecutive nights taken in the years 1965, 1983, 1986, 1991 and 1992 were investigated and found to show residuals of about 0.008\,mag. The standard deviation in $V$ for the differential photometric data of five nights observed in 1992, for example, amounted to $\pm$0.0069\,mag. The activity observed in $\beta$ Cas is lower than in 95 Vir (cf. Fig. \ref{activity}), which should consequently result in a smaller amplitude of the corresponding photometric variability (below 2.5\,mmag). Therefore, the available data are not sufficient for an unambiguous detection. According to \citet{Ammler2012}, who adopt a \vsini\ value of $\sim$70\kms, the projected rotational period is about 3\,d. New accurate photometric data are needed to search for variations on this time scale.

Another prominent case of chromospheric activity among $\delta$ Scuti pulsators is $\rho$ Pup (HD 67523, HR 3185), a high-amplitude (peak-to-peak amplitude\,$\approx$\,0.1\,mag) pulsator with a low \vsini\ value of about 8\kms\ \citep{Dravins1977}. The star has been extensively studied in the literature \citep{Yushchenko2015} and is the prototype of a subgroup of $\delta$ Scuti pulsators characterized by overabundances of iron-group and heavier chemical elements. However, the strength of the \ion{Ca}{ii} H\&K emission is very weak \citep{Mathias1997}, resulting in a $S$ index which is not distinguishable from inactive stars \citep{Pace2013}. We therefore do not expect any detectable variability due to the presence of spots in this star.

\section{Conclusion}

Employing data from Campaign 6 of the Kepler K2 mission, we have carried out a search for photometric variability in the fast rotating, chromospherically active early F-type star 95 Vir. Astrophysical parameters have been gleaned from the available literature, and the location of our target star has been investigated in the $M_{\rm Bol}$ versus $\log T_\mathrm{eff}$ diagram, from which information on evolutionary status was derived.

An analysis of the 3400 long-cadence measurements (time span of $\sim$78.9 days) procured from the K2 data archive at MAST yielded two main frequencies and several harmonics. We have attributed the dominant frequency, $f1$\,=\,9.53728\,d$^{-1}$, to $\delta$ Scuti pulsations and, drawing on literature information and the theoretical predictions of state-of-the-art pulsation models, discussed three possible interpretations of the low-frequency signal $f2$\,=\,1.07129\,d$^{-1}$: binarity, pulsation and rotational modulation. Current evidence favours an interpretation as a signature of the rotational period caused by the presence of cool star spots. This scenario is in line with the observed chromospheric activity in our target star.

We have briefly considered the cases of $\beta$ Cas and $\rho$ Pup, two $\delta$ Scuti pulsators which have been reported to exhibit chromospheric activity in the literature. A search for star spot-induced photometric variability in these objects might be of great interest, as well as an investigation of the interplay between chromospheric and pulsational activity.

\section*{Acknowledgements}
We thank the referee for helpful suggestions. This work was supported by the grant 7AMB17AT030 (M\v{S}MT). PW's work was financially supported by the Polish NCN grants 2015/17/B/ST9/02082. Calculations have been partly carried out using resources provided by Wroclaw Centre for Networking and Supercomputing (http://www.wcss.pl), grant No. 265.


\bibliographystyle{mnras}
\bibliography{95vir}

\bsp	
\label{lastpage}
\end{document}